\documentclass{jfm}

\usepackage{amsbsy,amssymb,amsmath,bm}
\usepackage{graphicx,subfigure}
\usepackage{tabularx,booktabs}
\usepackage[authoryear,round]{natbib}

\title[Responses in rotating flow]{Linear and nonlinear responses to harmonic force in rotating flow}
\author[X. Wei]{Xing Wei$^{1,2}$\thanks{Email: xing.wei@sjtu.edu.cn, xingwei@astro.princeton.edu}}
\affiliation{$^1$Institute of Natural Sciences and Department of Physics and Astronomy, Shanghai Jiao Tong University, Shanghai 200240, China \\ $^2$Princeton University Observatory, Princeton, NJ 08544, USA}
\date{?; revised ?; accepted ?.}

\begin{document}

\maketitle

\begin{abstract}
For understanding the dissipation in a rotating flow when resonance occurs, we study the rotating flow driven by the harmonic force in a periodic box. Both the linear and nonlinear regimes are studied. The various parameters such as the force amplitude $a$, the force frequency $\omega$, the force wavenumber $k$, and the Ekman number $E$ are investigated. In the linear regime, the dissipation at the resonant frequency scales as $E^{-1}k^{-2}$, and it is much stronger than the dissipation at the non-resonant frequencies on the large scales and at the low Ekman numbers. In the nonlinear regime, at the resonant frequency the effective dissipation (dissipation normalised with the square of force amplitude) is lower than in the linear regime and it decreases with the increasing force amplitude. This nonlinear suppression effect is significant near the resonant frequency but negligible far away from the resonant frequency. Opposite to the linear regime, in the nonlinear regime at the resonant frequency the lower Ekman number leads to the lower dissipation because of the stronger nonlinear effect. This work implies that the previous linear calculations overestimated the tidal dissipation, which is important for understanding the tides in stars and giant planets.

\end{abstract}

\section{Introduction}\label{sec:introduction}

Rotation plays an important role in the engineering, geophysical and astrophysical fluid motions. It induces inertial waves of which the Coriolis force acts as the restoring force \citep{greenspan}. The dispersion relationship for inertial wave is
\begin{equation}\label{eq:dispersion}
\sigma=\pm\frac{2\bm\Omega\cdot\bm k}{k},
\end{equation}
where $\sigma$ is the wave frequency, $\bm k$ the wave vector and $\bm\Omega$ the angular velocity of rotation. This expression shows that the frequency of inertial wave is in the range $|\sigma|\le2\Omega$. Inertial wave is dispersive and its group velocity is perpendicular to its phase velocity. Inertial wave carries energy and angular momentum in the interior of fluid and then dissipates through viscosity. Moreover, it has helical structure which favours the dynamo action for generating magnetic field, e.g. \citet{moffatt_kinematic,moffatt_dynamic,davidson,wei}, etc. 

In the geometry of an annular channel, the problem of inertial waves was studied by \citet{zhang_channel}, and in the spherical geometry it has been extensively studied, e.g. \citet{hollerbach1,rieutord,ogilvie_wave-attractors,tilgner_inertial}, etc. Because of the singularity of Poincar\'e equation, i.e. the governing equation of inertial waves, the inertial waves in the spherical geometry are spawn from the critical latitude and then propagate and reflect in the thin shear layers, i.e. the wave attractors \citep{ogilvie_wave-attractors}. In the Earth's fluid core, the inertial waves driven by precession are discussed by \citet{busse,kerswell_precession,tilgner_cavity,tilgner_shell,tilgner_review}, and \citet{zhang_spheroid}. 

Inertial waves can be excited by the tidal force in the planetary and stellar interiors. Tide exists widely in astronomical binary systems, e.g. Earth and Moon, giant planet and its satellite, host star and exoplanet, binary normal stars or white dwarfs, etc. In the binary system, one body (primary) is gravitationally perturbed by the other (companion) such that the primary deforms and produces the tidal bulge pointing to the companion. Tidal torque transfers angular momentum between the orbital motion and the rotational motion of binaries, and the dissipation in the planetary and stellar interiors plays an important role for the angular momentum transfer. Waves can be excited by the harmonic tidal force, i.e. the dynamical tide, and the dissipation of these waves is very efficient because of their small scales. Particularly, the inertial waves induced by the tidal force, i.e. the dynamical tide arising from rotation, are discussed by \citet{kerswell_tide}, \citet{goodman}, and \citet{ogilvie_review}. Tidally excited inertial waves dissipate through viscosity, and the tidal dissipation becomes very strong when the resonance occurs, namely the tidal frequency is close to the eigen-frequency of the inertial wave in the unforced rotating flow. There are infinite inertial eigen-modes in the spherical geometry and therefore the tidal resonance is prone to occur as long as the tidal frequency is less than twice of the rotation frequency. In the nonlinear regime, i.e. with the presence of nonlinear inertial force, it was pointed out by \citet{tilgner_inertial} that the inertial waves can generate zonal flow. In addition, the nonlinear wave-breaking has significant effect on the tidal dissipation \citep{goodman}. The nonlinear tidal flow was numerically studied by \citet{ogilvie_nonlinear} with the boundary radial flow method and by \citet{hollerbach2} with the body force method. Although the linear regime is extensively studied, the nonlinear regime is not well understood, e.g. the scaling laws of tidal dissipation versus Ekman number is unknown (\citet{ogilvie_nonlinear} studied a little about the scaling laws, see figure 4, but their study is at the non-resonant frequencies and the nonlinear dissipation at the resonant frequency is still unknown). In our study we will focus on the nonlinear effect at both the resonant and non-resonant frequencies.

In this short paper, we will study the rotating flow driven by the harmonic force in a periodic box, i.e. a cubic box with the periodic boundary condition, which is a toy model for a small piece of region in a container (for engineering) or a star (for astrophysics). Not as in the spherical geometry, the inertial waves in a periodic box do not reflect but propagate forward, and hence do not focus in the thin shear layers. In section \S\ref{sec:equations} the equations are given. In section \S\ref{sec:linear} the linear regime is analytically studied and the dissipation at the resonant frequency is derived. In section \S\ref{sec:nonlinear} the nonlinear regime is numerically studied and compared to the linear regime. In section \S\ref{sec:conclusion} a brief summary is given.

\section{Equations}\label{sec:equations}

We study the rotating flow of an incompressible fluid in a periodic box with its size $2\pi l$. We use the Cartesian coordinate system $(x,y,z)$ and the uniform rotation is imposed in the $z$ direction. In the frame rotating at the angular velocity $\bm\Omega=\Omega\hat{\bm z}$, the dimensionless Navier-Stoke equation of fluid motion reads
\begin{equation}\label{eq:ns}
\frac{\partial\bm u}{\partial t}+\bm u\cdot\bm\nabla\bm u=-\bm\nabla p+E\nabla^2\bm u+2\bm u\times\hat{\bm z}+\bm f,
\end{equation}
where length is normalised with $l$, time with the inverse of rotation frequency $\Omega^{-1}$ and velocity with $\Omega l$. The Ekman number $E=\nu/(\Omega l^2)$, where $\nu$ is viscosity, measures the ratio of rotational time scale to viscous time scale.

The driving force is assumed to be a single traveling wave, i.e.
\begin{equation}\label{eq:force}
\bm f=\Re\{\hat{\bm f}e^{i(\bm k\cdot\bm x-\omega t)}\},
\end{equation}
where $\hat{\bm f}$ is the complex force amplitude, $\bm k$ the force wavevector, $\omega$ the force frequency, and $\Re$ denotes taking the real part. The tidal force exerted by the companion on the primary is the difference between the force on any point and the force at the centre of the primary, and the tidal potential is the superposition of spherical harmonics with the time dependence on the Doppler-shifted frequency \citep{souchay,ogilvie_review}. Although the total tidal force is curl-free, its contribution to the dynamical tide is vortical because of the very slow equilibrium tide, see the details in Appendix B of \citet{ogilvie_wave-attractors}. Briefly speaking, the incompressible equilibrium tide varies slowly and does not satisfy the hydrostatic balance such that the residual is a vortical force that can drive the dynamical tide, e.g. the inertial waves in rotating fluid. In our simplified model, $\bm f$ corresponds to the force responsible for the dynamical tide and it is not curl-free. On the other hand, to have the dynamical effect on flow, the driving force $\bm f$ should not be curl-free (if it is curl-free then it can be absorbed into pressure gradient to act as the additional pressure). For the simplicity to derive the solution in the linear regime (see section \S\ref{sec:linear}), we assume it to be a helical force, i.e. $\bm\nabla\times\bm f=k\bm f$ where $k=|\bm k|$ is the force wavenumber. One may argue that the helical force is too artificial. Here we give more explanation. Any vector field can be decomposed to the curl-free part and the divergence-free part, i.e. the Helmholtz decomposition. Moreover, the divergence-free part can be decomposed into helical modes, see \citet{waleffe}. Back to the driving force $\bm f$, the curl-free part can be absorbed into the pressure gradient and the divergence-free part can be expressed as the superposition of helical forces. This is the reason that we use the helical force for the study of tidal waves. In the spectral space the helical force satisfies
\begin{equation}\label{eq:helical}
i\bm k\times\hat{\bm f}=k\hat{\bm f}.
\end{equation}
Equation \eqref{eq:helical} is degenerate (i.e. only two components are independent) and yields
\begin{equation}\label{eq:degenerate}
\frac{\hat f_y}{\hat f_x}=\frac{-k_xk_y+ikk_z}{k_y^2+k_z^2}, \;
\frac{\hat f_z}{\hat f_y}=\frac{-k_yk_z+ikk_x}{k_z^2+k_x^2}, \;
\frac{\hat f_x}{\hat f_z}=\frac{-k_zk_x+ikk_y}{k_x^2+k_y^2}.
\end{equation}
We denote the module of the complex force amplitude by $a$, i.e. 
\begin{equation}\label{eq:amplitude=a}
|\hat{\bm f}|=\sqrt{|\hat f_x|^2+|\hat f_y|^2+|\hat f_z|^2}=a.
\end{equation}
Equations \eqref{eq:degenerate} and \eqref{eq:amplitude=a} then combine to yield
\begin{equation}\label{eq:amplitude}
|\hat f_x|=\frac{\sqrt{k_y^2+k_z^2}}{\sqrt{2}k}a, \; 
|\hat f_y|=\frac{\sqrt{k_z^2+k_x^2}}{\sqrt{2}k}a, \; 
|\hat f_z|=\frac{\sqrt{k_x^2+k_y^2}}{\sqrt{2}k}a,
\end{equation}
and in addition, the arguments of $\hat f_y/\hat f_x$ and $\hat f_z/\hat f_x$ are, respectively,
\begin{equation}\label{eq:argument}
\pi-\arccos\frac{k_xk_y}{\sqrt{(k_y^2+k_z^2)(k_z^2+k_x^2)}}, \;
\pi+\arccos\frac{k_zk_x}{\sqrt{(k_x^2+k_y^2)(k_y^2+k_z^2)}}.
\end{equation}
The arguments of $\hat f_x$, $\hat f_y$ and $\hat f_z$ themselves are insignificant for the volume integral of energy and dissipation, but the differences between them (i.e. phase shifts) do matter, and without loss of generality the argument of $\hat f_x$ is given to be $0$. Thus, equations \eqref{eq:amplitude} and \eqref{eq:argument} give the three components of the complex amplitude $\hat{\bm f}$, and equation \eqref{eq:force} gives the driving force in the physical space. 

The output that we are concerned with is the volume-averaged dissipation. With the periodic boundary condition, it is proportional to enstrophy, i.e.
\begin{equation}\label{eq:dissipation-physical}
D=E\frac{1}{V}\int_V|\bm\nabla\times\bm u|^2dV=\frac{E}{2}|i\bm k\times\hat{\bm u}|^2.
\end{equation}

The numerical calculations are carried out with the pseudo-spectral code using fast Fourier transform. The resolution is checked with two methods. One is to double the resolution until the total energy and enstrophy have no noticeable change. The other is to see whether the energy and enstrophy spectra decay by sufficient (say, more than 10) magnitudes. In our moderate parameters regime, the resolutions as high as $128^3$ are used.

\section{Linear regime}\label{sec:linear}

In the absence of the nonlinear inertial force, we can analytically solve the linearised Navier-Stokes equation
\begin{equation}\label{eq:linear-ns}
\frac{\partial\bm u}{\partial t}=-\bm\nabla p+E\nabla^2\bm u+2\bm u\times\hat{\bm z}+\bm f.
\end{equation}
Because the driving force is a single harmonic (equation \eqref{eq:force}), the solution to the linear equation is assumed to be $\bm u=\Re\{\hat{\bm u}e^{i(\bm k\cdot\bm x-\omega t)}\}$ and $p=\Re\{\hat pe^{i(\bm k\cdot\bm x-\omega t)}\}$. Substitution into \eqref{eq:linear-ns} yields
\begin{equation}
-i\omega\hat{\bm u}=-i\bm k \hat p-Ek^2\hat{\bm u}+2\hat{\bm u}\times\hat{\bm z}+\hat{\bm f}.
\end{equation}
By performing $i\bm k\times$ on the above equation to eliminate pressure and applying $i\bm k\cdot\hat{\bm u}=0$ (incompressible flow) and $i\bm k\times\hat{\bm f}=k\hat{\bm f}$ (helical force), we derive
\begin{equation}
(\omega+iEk^2)\bm k\times\hat{\bm u}=2ik_z\hat{\bm u}+k\hat{\bm f}.
\end{equation}
Performing again $i\bm k\times$ on the above equation yields
\begin{equation}
k^2(\omega+iEk^2)\hat{\bm u}=-2ik_z\bm k\times\hat{\bm u}+ik^2\hat{\bm f}.
\end{equation}
Combining the above two equations to eliminate $i\bm k\times\hat{\bm u}$, we are led to
\begin{equation}\label{eq:resonance}
\left[(2k_z)^2-k^2(\omega+iEk^2)^2\right]\hat{\bm u}=ik\left[2k_z-k(\omega+iEk^2)\right]\hat{\bm f}.
\end{equation}
When the driving force is absent ($\hat{\bm f}=\bm 0$) and viscosity vanishes ($E=0$), equation \eqref{eq:resonance} reduces to the dispersion relationship for inertial wave, i.e. $\sigma=\pm2k_z/k$ (where the eigen-frequency is denoted by $\sigma$). Because the factor $\left[2k_z-k(\omega+iEk^2)\right]$ can never be zero due to the phase shift caused by viscosity, it is cancelled and we derive
\begin{equation}\label{eq:solution}
\hat{\bm u}=\frac{ik\hat{\bm f}}{2k_z+k(\omega+iEk^2)}.
\end{equation}
Equation \eqref{eq:solution} is the solution to the linearised Navier-Stokes equation \eqref{eq:linear-ns}. By virtue of equation \eqref{eq:dissipation-physical}, dissipation can be calculated as
\begin{equation}\label{eq:dissipation-spectral}
D=\frac{E}{2}|i\bm k\times\hat{\bm u}|^2=\frac{E}{2}\frac{a^2k^4}{|2k_z+k(\omega+iEk^2)|^2}.
\end{equation}

The resonance occurs when the linear response \eqref{eq:solution} is singular with the neglect of viscosity. In the presence of viscosity, the linear response \eqref{eq:solution} becomes very strong at the frequency
\begin{equation}\label{eq:resonant-frequency}
\omega=-\frac{2k_z}{k},
\end{equation}
which is called the \emph{resonant frequency}. Substitution of \eqref{eq:resonant-frequency} into \eqref{eq:solution} leads to the response at the resonant frequency
\begin{equation}
\hat{\bm u}=\frac{1}{Ek^2}\hat{\bm f}.
\end{equation}
Substitution of \eqref{eq:resonant-frequency} into \eqref{eq:dissipation-spectral} leads to the dissipation at the resonance frequency
\begin{equation}
D=\frac{a^2}{2Ek^2}.
\end{equation}
Therefore, the dissipation at the resonant frequency scales as
\begin{equation}\label{eq:scaling-D}
D\propto E^{-1}k^{-2}.
\end{equation}

\begin{figure}
\centering
\includegraphics[scale=0.3]{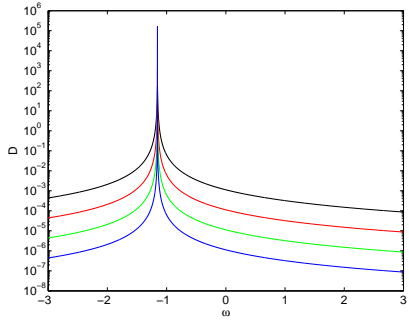}
\caption{The investigation of the force frequency in the linear regime. The dissipation $D$ versus the force frequency $\omega$. $a=1$. $k_x=k_y=k_z=1$. Black, red, green and blue lines denote respectively $E=10^{-3}$, $10^{-4}$, $10^{-5}$ and $10^{-6}$.}\label{fig:lin1}
\end{figure}

According to \eqref{eq:dissipation-spectral}, we can calculate the dissipation of the linear response. For the linear response we fix $a=1$. Firstly we study the effect of the force frequency on the linear response. We calculate at the four Ekman numbers $10^{-3}$, $10^{-4}$, $10^{-5}$ and $10^{-6}$, and at the fixed wavenumbers $k_x=k_y=k_z=1$. Figure \ref{fig:lin1} shows the dissipation versus the force frequency. It is verified that the dissipation has a sharp peak at the resonant frequency $\omega=-2k_z/k=-2/\sqrt{3}\approx-1.1547$. It also indicates that a lower Ekman number corresponds to a higher peak, which is consistent with \eqref{eq:scaling-D}. We pick out two representative frequencies. One frequency is $\omega=-1.16$ which is considered to be \emph{near the resonant frequency} (e.g. at $E=10^{-3}$ the dissipation at $\omega=-1.16$ is $24\%$ of the dissipation at the resonant frequency). The other is $\omega=-1.2$ which is considered to be \emph{far away from the resonant frequency} (e.g. at $E=10^{-3}$ the dissipation at $\omega=-1.2$ is $0.44\%$ of the dissipation at the resonant frequency). In the next calculations throughout this paper, we will often use these two representative frequencies.

\begin{figure}
\centering
\includegraphics[scale=0.3]{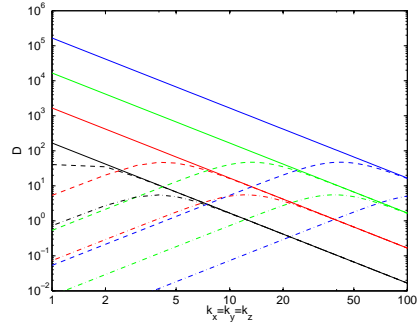}
\caption{The investigation of the force wavenumber in the linear regime. The dissipation $D$ versus the force wavenumber $k_x=k_y=k_z$. $a=1$. Black, red, green and blue lines denote respectively $E=10^{-3}$, $10^{-4}$, $10^{-5}$ and $10^{-6}$. Solid lines denote the resonant frequency $\omega=-2/\sqrt{3}$, dashed lines $\omega=-1.16$ near the resonant frequency, and dahsed dotted lines $\omega=-1.2$ far away from the resonant frequency.}\label{fig:lin2}
\end{figure}

Next we study the effect of the force wavenumber on the linear response. We calculate at the four Ekman numbers as in the last paragraph and at the three frequencies, namely the resonant frequency $\omega=-2/\sqrt{3}$, $\omega=-1.16$ near the resonant frequency, and $\omega=-1.2$ far away from the resonant frequency. The resonant frequency depends on the orientation of the wave vector. To keep the resonant frequency fixed, we keep $k_x=k_y=k_z$. Figure \ref{fig:lin2} shows the dissipation versus the force wavenumber. It is verified that the dissipation scales as $D\propto k^{-2}$ at the resonant frequency, as predicted by \eqref{eq:scaling-D}. When the force wavenumber is sufficiently large, the dissipation at the other frequencies converges to the dissipation at the resonant frequency. This suggests that the dissipation at the resonant frequency is much stronger than the dissipation at the non-resonant frequency \emph{on the large scales}.

\begin{figure}
\centering
\includegraphics[scale=0.3]{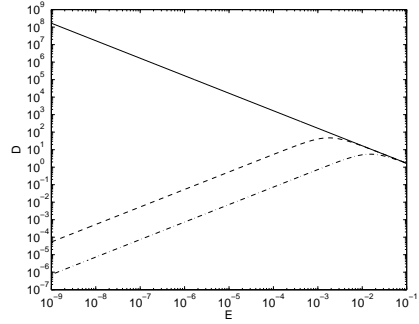}
\caption{The investigation of the Ekman number in the linear regime. The dissipation $D$ versus the Ekman number $E$. $a=1$. $k_x=k_y=k_z=1$. Solid line denotes $\omega=-2/\sqrt{3}$, dashed line $\omega=-1.16$ near the resonant frequency, and dash-dotted line $\omega=-1.2$ far away from the resonant frequency.}\label{fig:lin3}
\end{figure}

We then study the effect of the Ekman number on the linear response. Since we know that the resonance has the striking effect on the large scales, we fix $k_x=k_y=k_z=1$. Figure \ref{fig:lin3} shows the dissipation versus $E$ at the three frequencies. It indicates that the dissipation at the resonant frequency scales as $D\propto E^{-1}$, as predicted by \eqref{eq:scaling-D}. The dissipation at the other frequencies scales as $D\propto E$ in the regime of low Ekman number and converges to the dissipation at the resonant frequency in the regime of high Ekman number. This suggests that the dissipation at the resonant frequency is much stronger than the dissipation at the non-resonant frequencies \emph{at the low Ekman numbers}.

In addition, we also carried out the direct numerical calculations of equation \eqref{eq:linear-ns} and the relative error of the numerical calculations compared to the analytical results is within $0.01\%$.

\section{Nonlinear regime}\label{sec:nonlinear}

\begin{figure}
\centering
\subfigure[]{\includegraphics[scale=0.23]{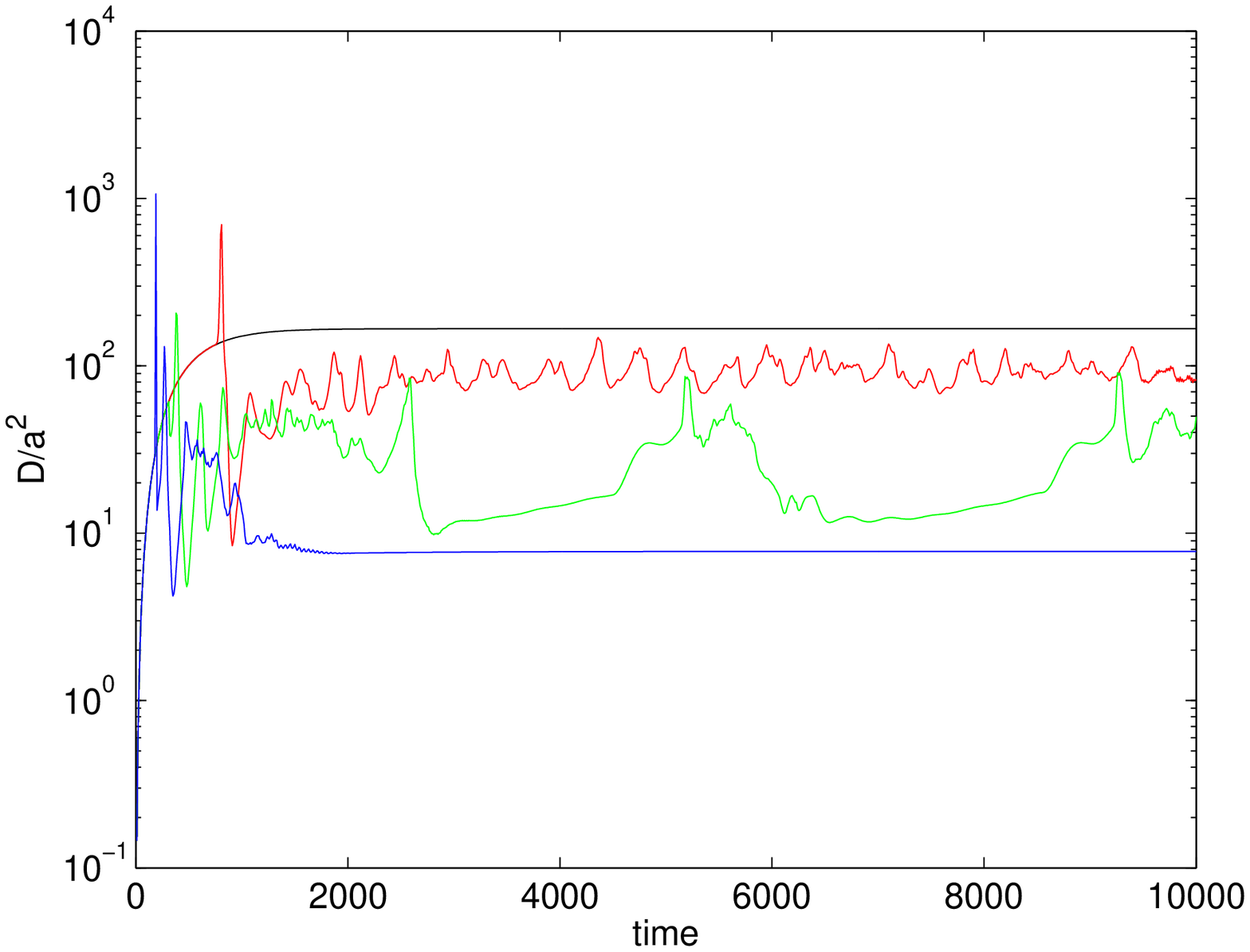}\label{fig:non1}}
\subfigure[]{\includegraphics[scale=0.23]{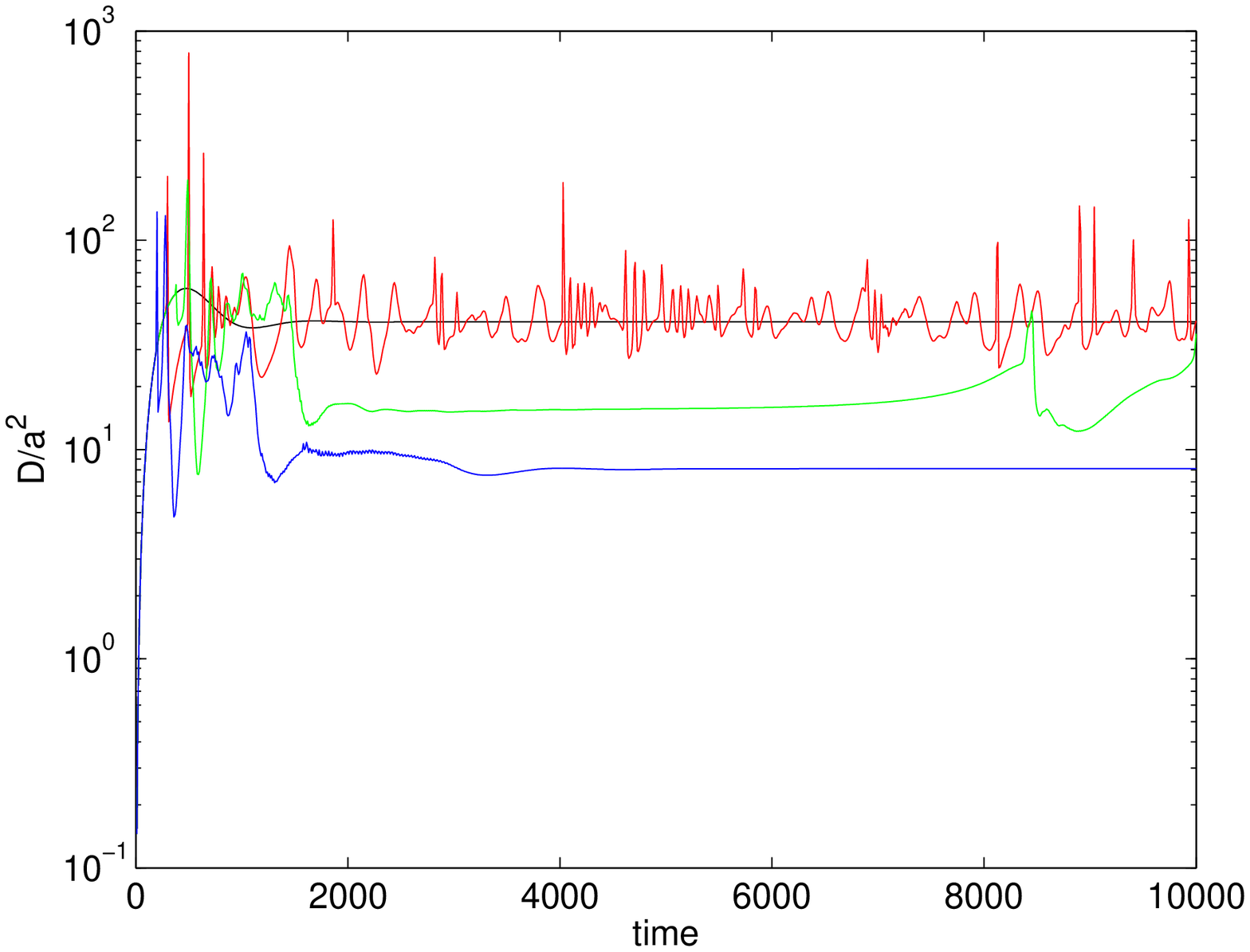}\label{fig:non2}}
\subfigure[]{\includegraphics[scale=0.23]{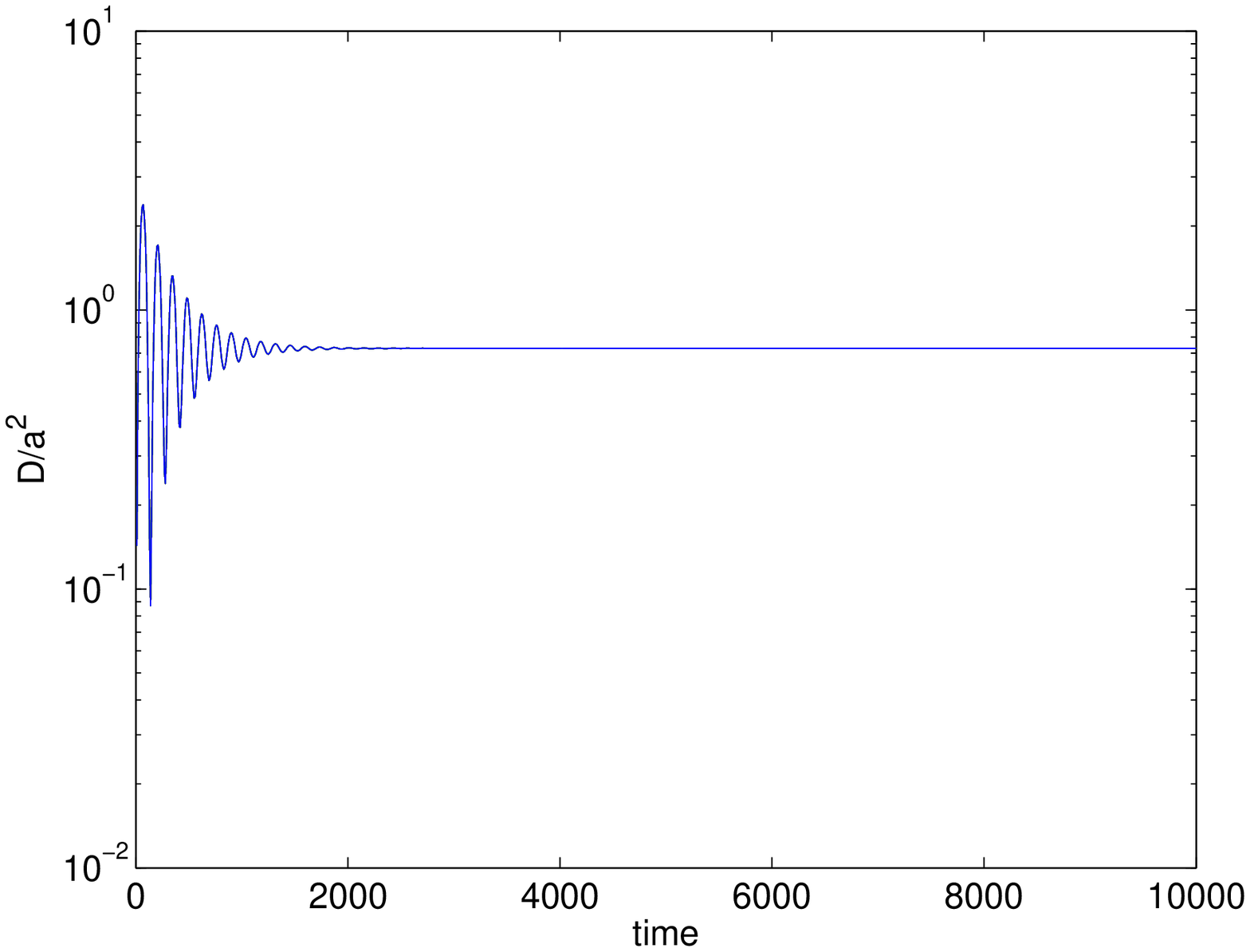}\label{fig:non3}}
\caption{The investigation of the force amplitude in the nonlinear regime. The normalised dissipation $D/a^2$ versus time at $E=1\times10^{-3}$ with $k_x=k_y=k_z=1$. (a) The resonant frequency $\omega=-2/\sqrt{3}$. (b) $\omega=-1.16$ near the resonant frequency. (c) $\omega=-1.2$ far away from the resonant frequency. Black lines denote the results in the linear regime. Red, green and blue lines denote the nonlinear regime with respectively $a=1\times10^{-3}$, $2\times10^{-3}$, and $3\times10^{-3}$.}\label{fig:amplitude}
\end{figure}

We now numerically study the nonlinear regime, i.e. solving equation \eqref{eq:ns}. Firstly we study the nonlinear effect due to the force amplitude. We gradually increase the force amplitude $a$ from $1\times10^{-3}$ to $2\times10^{-3}$ and then to $3\times10^{-3}$. The stronger force drives the stronger flow and hence the higher dissipation, and so we normalise the dissipation with $a^2$, which reflects the nonlinear effect on the dissipation and we call $D/a^2$ the \emph{effective dissipation}. Figure \ref{fig:amplitude} shows the effective dissipation $D/a^2$ versus time for the three amplitudes, $1\times10^{-3}$, $2\times10^{-3}$ and $3\times10^{-3}$, at $E=1\times10^{-3}$. The three different frequencies are studied, i.e. the resonant frequency $\omega=-2/\sqrt{3}\approx-1.1547$, $\omega=-1.16$ near the resonant frequency, and $\omega=-1.2$ far away from the resonant frequency. The linear results are also shown in the figure for comparison with the nonlinear results. Figure \ref{fig:non1} shows that at the resonant frequency the effective dissipation is lower than in the linear regime and it decreases with the increasing force amplitude, namely the stronger nonlinearity has a greater suppression effect on the effective dissipation. It is interesting that the rapid fluctuations occur with $a=1\times10^{-3}$ while the linear result has no such fluctuations. Evidently these fluctuations arise from some instabilities caused by the nonlinearity. They become less frequent with $a=2\times10^{-3}$ and vanish with $a=3\times10^{-3}$. Figure \ref{fig:non2} shows that near the resonant frequency this nonlinear suppression for dissipation still exists but becomes weaker, e.g. the black line (the linear regime) and the time average of the unsteady red line (the weakest nonlinear regime) almost overlap. Figure \ref{fig:non3} shows that far away from the resonant frequency the nonlinear suppression is absent, namely the black, red, green and blue lines completely overlap. This is because the flow amplitude at the frequency far away from the resonant frequency is too small to have the strong nonlinear effect. In summary, the nonlinear effect suppresses the effective dissipation at the resonant frequency and this suppression effect is still significant near the resonant frequency but negligible far away from the resonant frequency.

To better understand the nonlinear effect on the dissipation, we investigate the dependence of the dissipation on the Rossby number, which measures the relative strength of the inertial force and the Coriolis force. The Rossby number is defined as $Ro=U/(l\Omega)$ where $U$ is the characteristic velocity. We take $U$ to be the square root of the volume-averaged kinetic energy, and under our normalisation the Rossby number is exactly the dimensionless $U$, i.e. 
\begin{equation}
Ro=\sqrt{\frac{1}{V}\int u^2dV}
\end{equation}
For the fluctuating flows, we take the time average in the statistically steady stage (after the initial transient growth stage) to calculate $Ro$. Table \ref{tab:non1} shows $Ro$ and $D/a^2$ versus $a$ at the three frequencies. It indicates that at the fixed frequency the larger force amplitude leads to the stronger nonlinearity ($Ro$) and hence the weaker $D/a^2$. The last row shows the ratio of the nonlinear dissipation to the corresponding linear dissipation, which clearly reveals that the nonlinear suppression is very strong at the resonant frequency for the large force amplitude. At the resonant frequency, the stronger force amplitude leads to the lower ratio. When the force frequency departs farther away from the resonant frequency, the nonlinearity becomes weaker.

\begin{table}
\centering
\begin{tabularx}{\textwidth}{c|lll|lll|lll}
$\omega$ & \multicolumn{3}{c|}{$-2/\sqrt{3}$} & \multicolumn{3}{c|}{$-1.16$} & \multicolumn{3}{c}{$-1.2$} \\
\hline
$a$ & 1e-3 & 2e-3 & 3e-3 & 1e-3 & 2e-3 & 3e-3 & 1e-3 & 2e-3 & 3e-3 \\
$Ro$    & 1.59e-1 & 1.78e-1 & 2.08e-1 & 9.19e-2 & 1.54e-1 & 1.87e-1 & 1.56e-2 & 3.12e-2 & 4.68e-2 \\
$D/a^2$ & 9.39e1  & 2.65e1  & 7.79e0  & 4.33e1  & 1.78e1  & 8.10e0  & 7.29e-1 & 7.29e-1 & 7.29e-1 \\
$D/D_{\rm lin}$ & 0.563 & 0.159 & 0.047 & 1.070 & 0.441 & 0.200 & 1.001 & 1.001 & 1.001
\end{tabularx}
\caption{The nonlinear regime. The time-averaged $Ro$, $D/a^2$ and $D/D_{\rm lin}$ versus $\omega$ and $a$ in figure \ref{fig:amplitude}.}\label{tab:non1}
\end{table}

\begin{figure}
\centering
\subfigure[]{\includegraphics[scale=0.6]{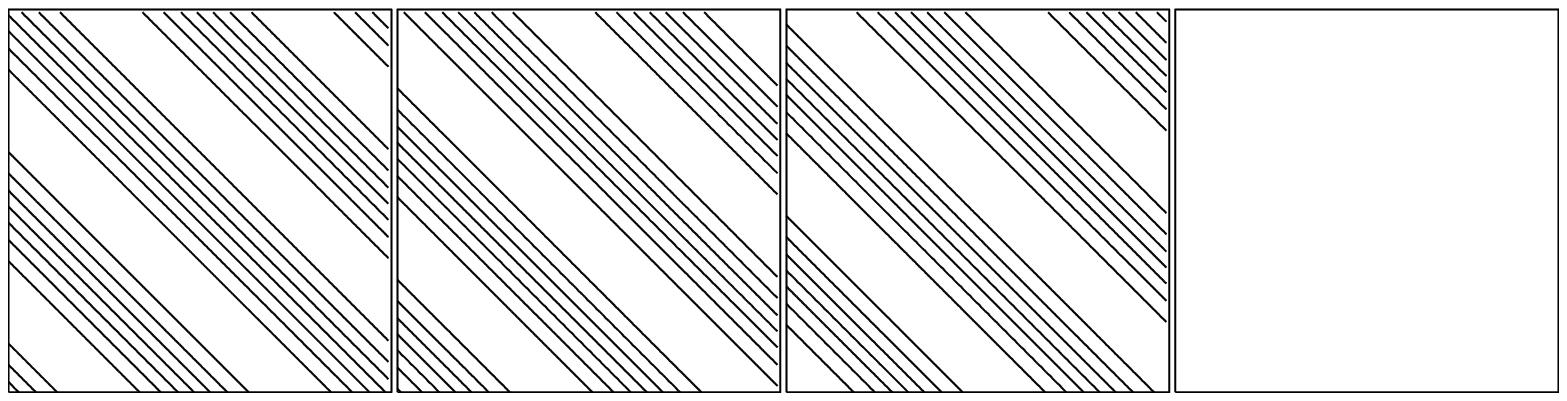}\label{fig:con0}}
\subfigure[]{\includegraphics[scale=0.6]{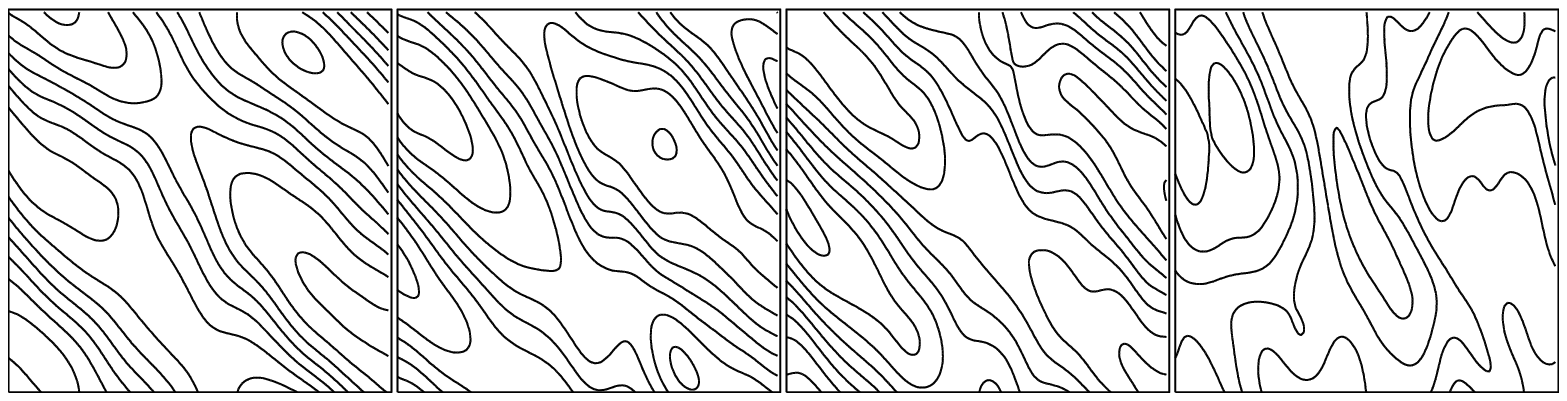}\label{fig:con1}}
\subfigure[]{\includegraphics[scale=0.6]{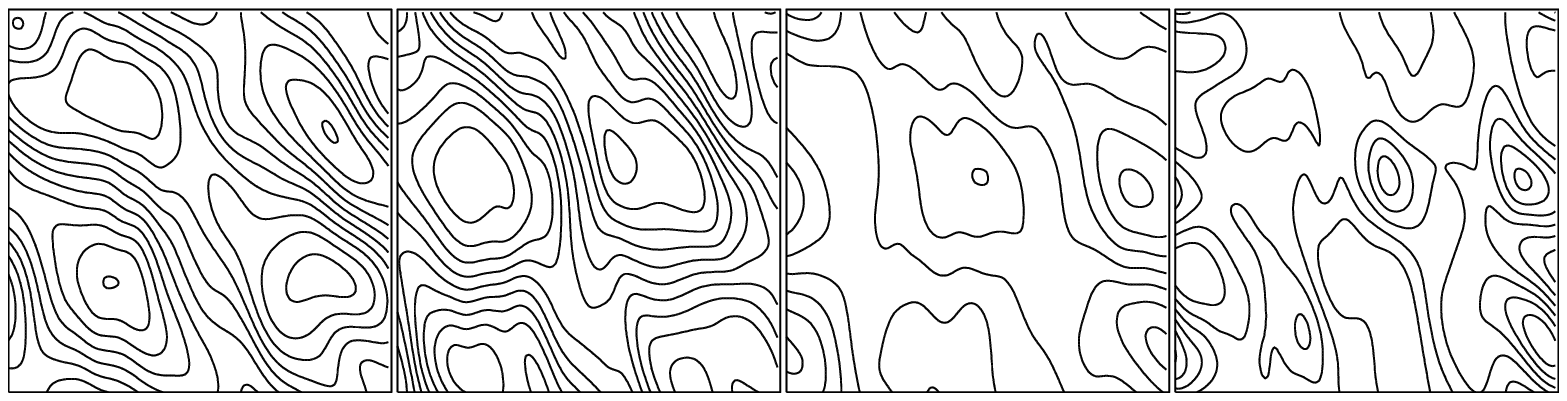}\label{fig:con2}}
\subfigure[]{\includegraphics[scale=0.6]{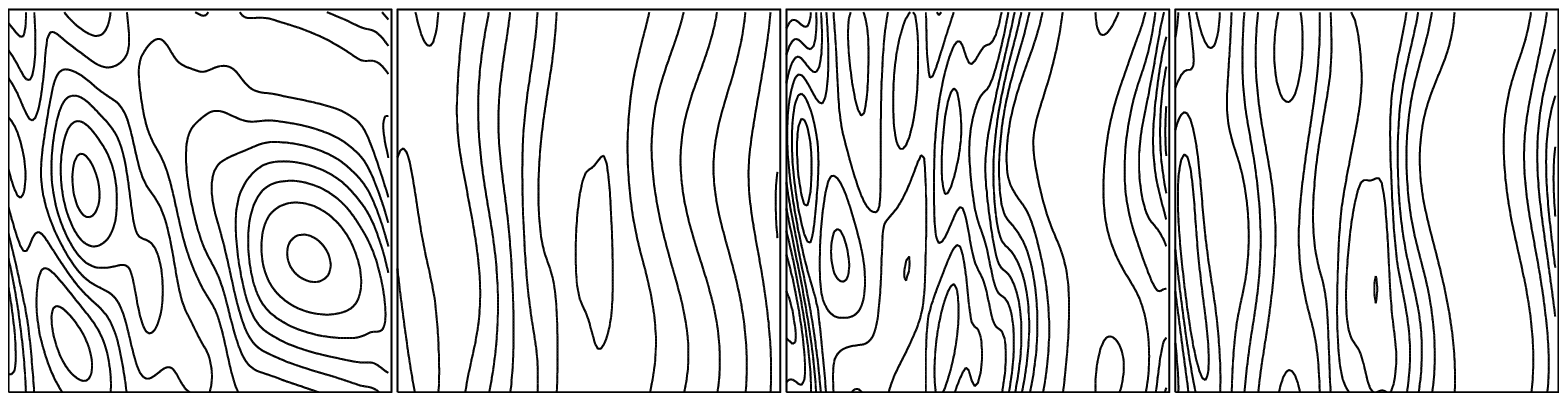}\label{fig:con3}}
\caption{The contours of velocity in the $x-z$ plane at $y=0$ for the snapshot at time=10000. In each subfigure the four panels from left to right are contours of respectively $u_1$, $u_2$, $u_3$ and kinetic energy. At the resonant frequency $\omega=-2/\sqrt{3}$ corresponding to figure \ref{fig:non1}. (a) Linear response with $a=1\times10^{-3}$ (the kinetic energy is a constant). (b) Nonlinear response with $a=1\times10^{-3}$. (c) Nonlinear response with $a=2\times10^{-3}$. (d) Nonlinear response with $a=3\times10^{-3}$.}\label{fig:contour}
\end{figure}

Figure \ref{fig:contour} shows the velocity structure at the resonant frequency $\omega=-2/\sqrt{3}$. With $a=1\times10^{-3}$ the main flow structure has the basic flow of ($k_x=k_y=k_z=1$) (figure \ref{fig:con0}) and the instabilities on top of the basic flow (figure \ref{fig:con1}), with $a=2\times10^{-3}$ the structure of basic flow alters due to the strong nonlinearity (figure \ref{fig:con2}), and at $a=3\times10^{-3}$ the structure of basic flow completely disappears and the z-independent structure emerges (figure \ref{fig:con3}).

\begin{figure}
\centering
\subfigure[]{\includegraphics[scale=0.23]{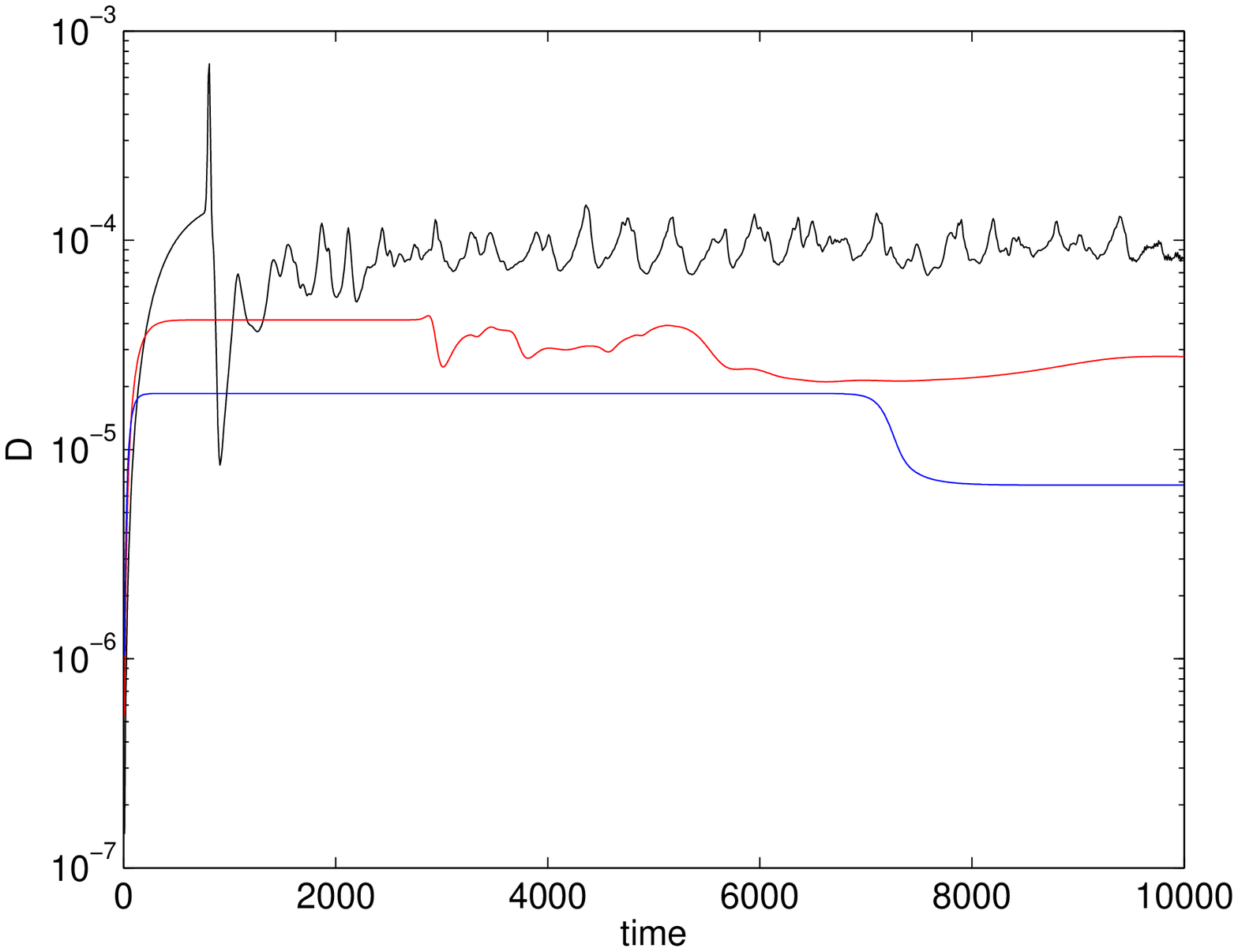}\label{fig:non4}}
\subfigure[]{\includegraphics[scale=0.23]{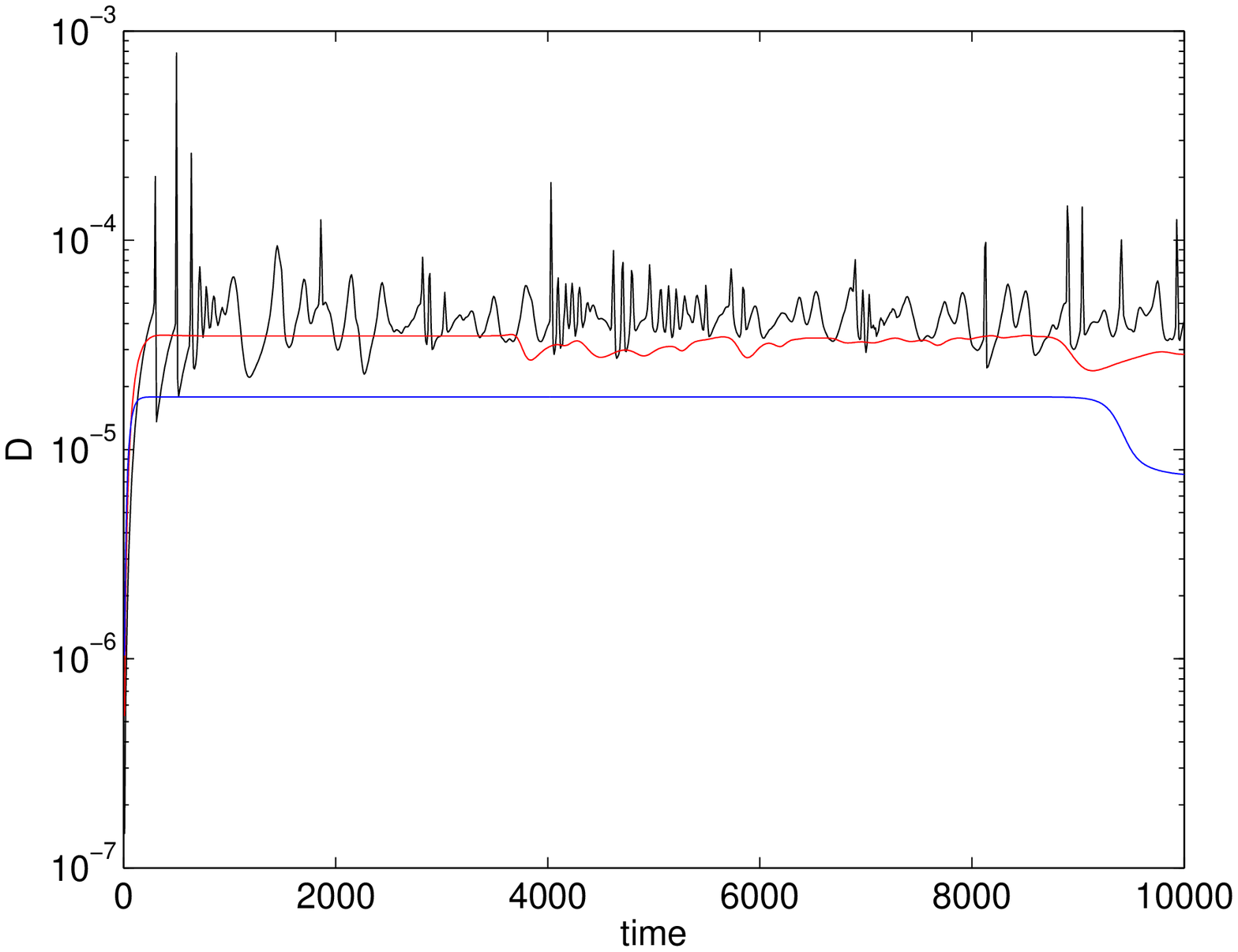}\label{fig:non5}}
\subfigure[]{\includegraphics[scale=0.23]{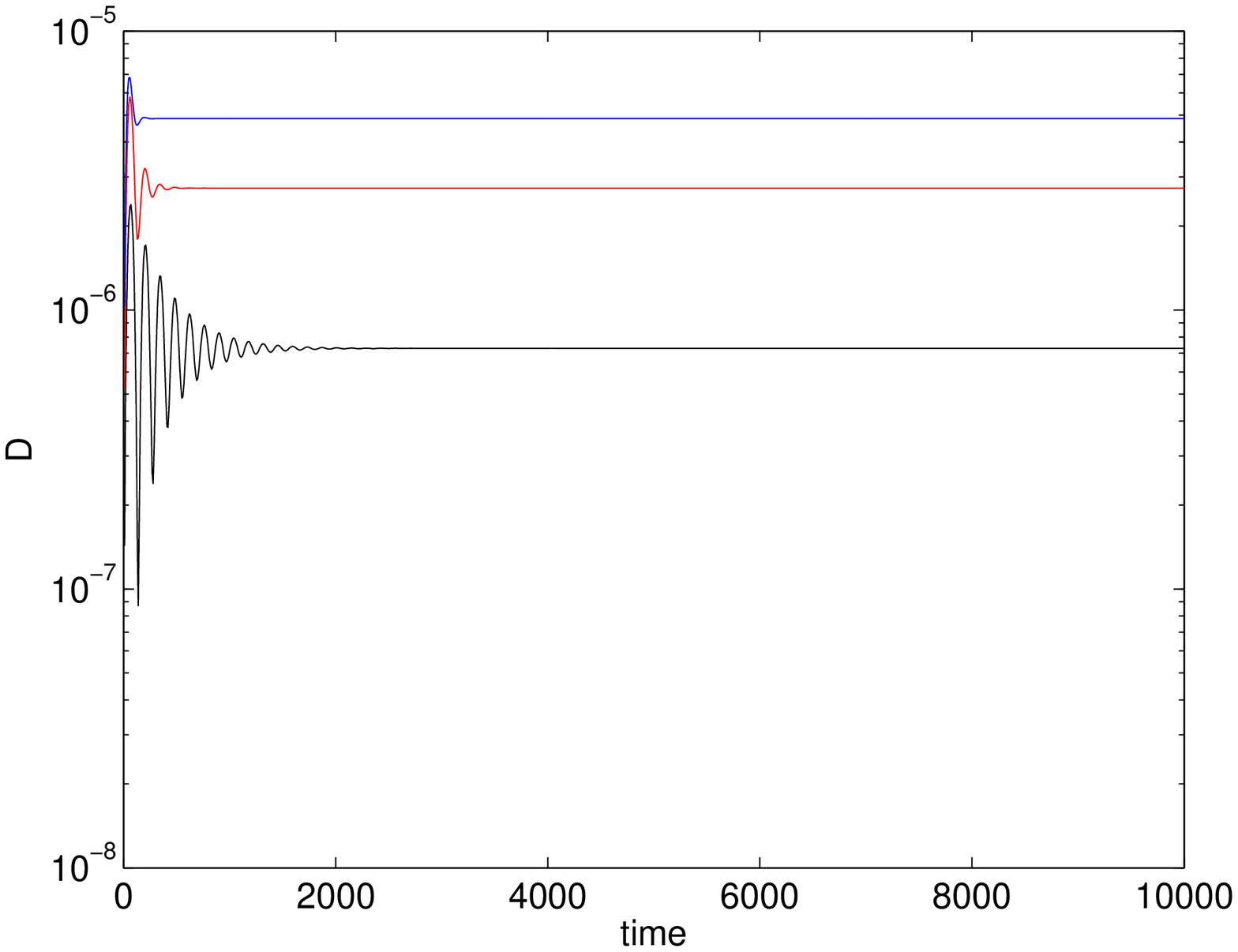}\label{fig:non6}}
\caption{The investigation of the force wavenumber in the nonlinear regime. The dissipation $D$ versus time at $E=1\times10^{-3}$ with $a=1\times10^{-3}$. (a) The resonant frequency $\omega=-2/\sqrt{3}$. (b) $\omega=-1.16$ near the resonant frequency. (c) $\omega=-1.2$ far away from the resonant frequency. Black, red and blue lines denote the nonlinear regime with respectively $k_x=k_y=k_z=1$, $2$ and $3$.}\label{fig:wavenumber}
\end{figure}

Next we investigate the force wavenumber. In the linear regime, the dissipation scales as $k^{-2}$ at the resonant frequency and it is much stronger than the dissipation at the non-resonant frequencies on the large scales, see figure \ref{fig:lin2}. In the nonlinear regime, figure \ref{fig:wavenumber} shows the dissipation versus time for the three wavenumbers, $k_x=k_y=k_z=1$, $2$ and $3$, with the three frequencies as in figure \ref{fig:amplitude}. At the resonant frequency (figure \ref{fig:non4}), the higher wavenumber leads to the lower dissipation, which is consistent with the prediction in the linear regime (the solid black line in figure \ref{fig:lin2}). At the frequency near the resonant frequency (figure \ref{fig:non5}), the higher wavenumber leads to the lower dissipation, which is also consistent with the prediction in the linear regime (the dashed black line in figure \ref{fig:lin2}). At the frequency far away from the resonant frequency (figure \ref{fig:non6}), the higher wavenumber leads to the higher dissipation, which is again consistent with the prediction in the linear regime (the dash-dotted black line in figure \ref{fig:lin2}).

\begin{figure}
\centering
\subfigure[]{\includegraphics[scale=0.23]{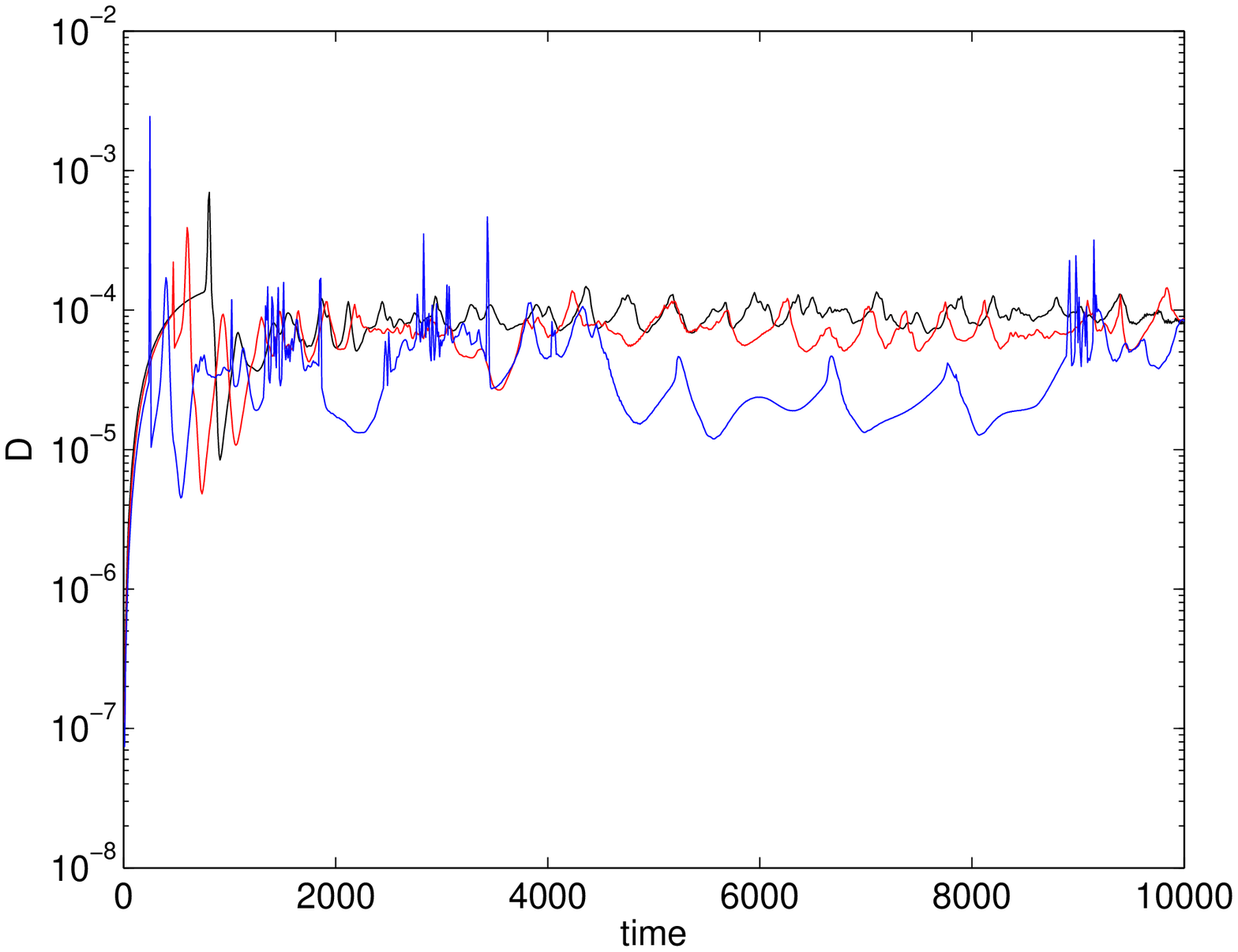}\label{fig:non7}}
\subfigure[]{\includegraphics[scale=0.23]{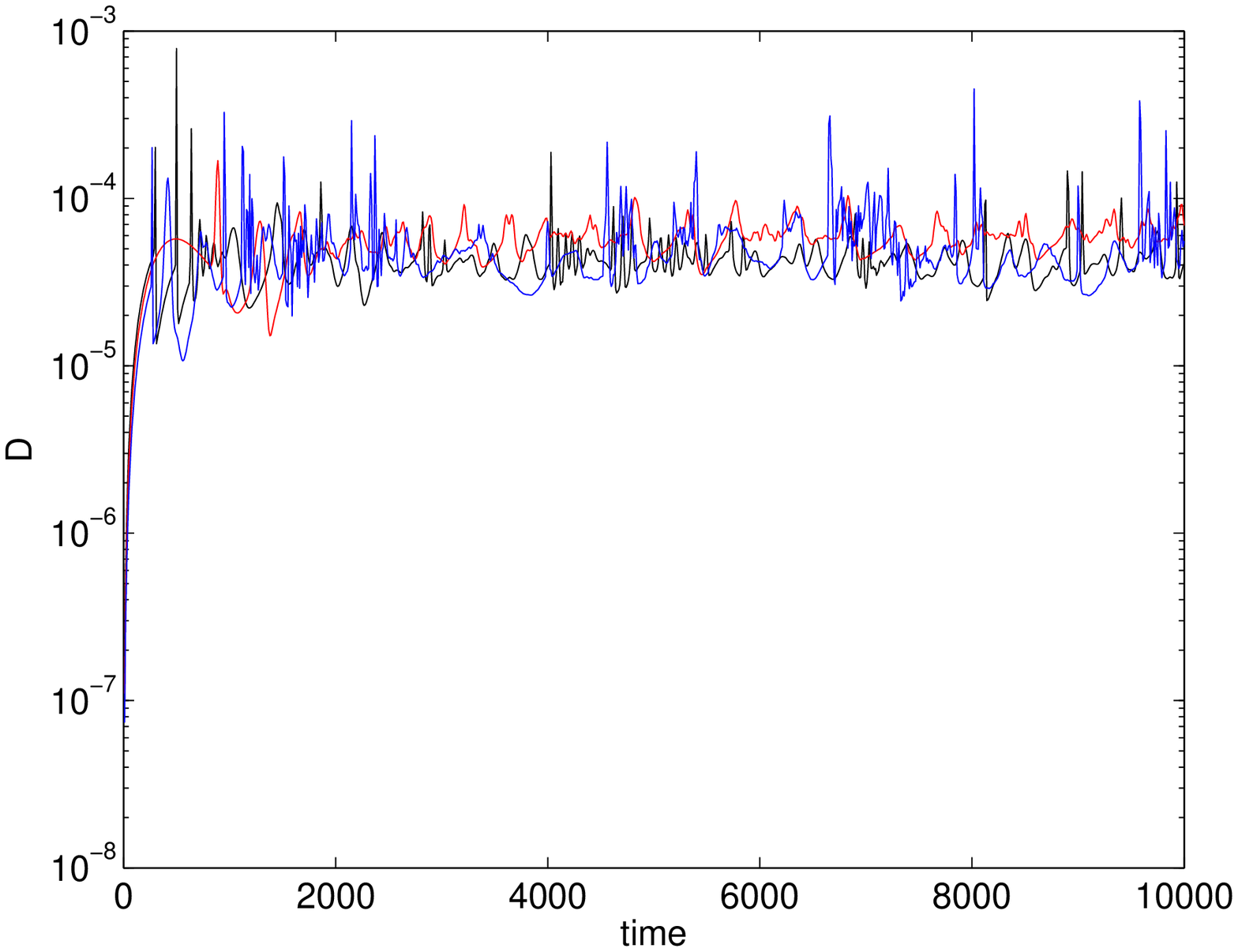}\label{fig:non8}}
\subfigure[]{\includegraphics[scale=0.23]{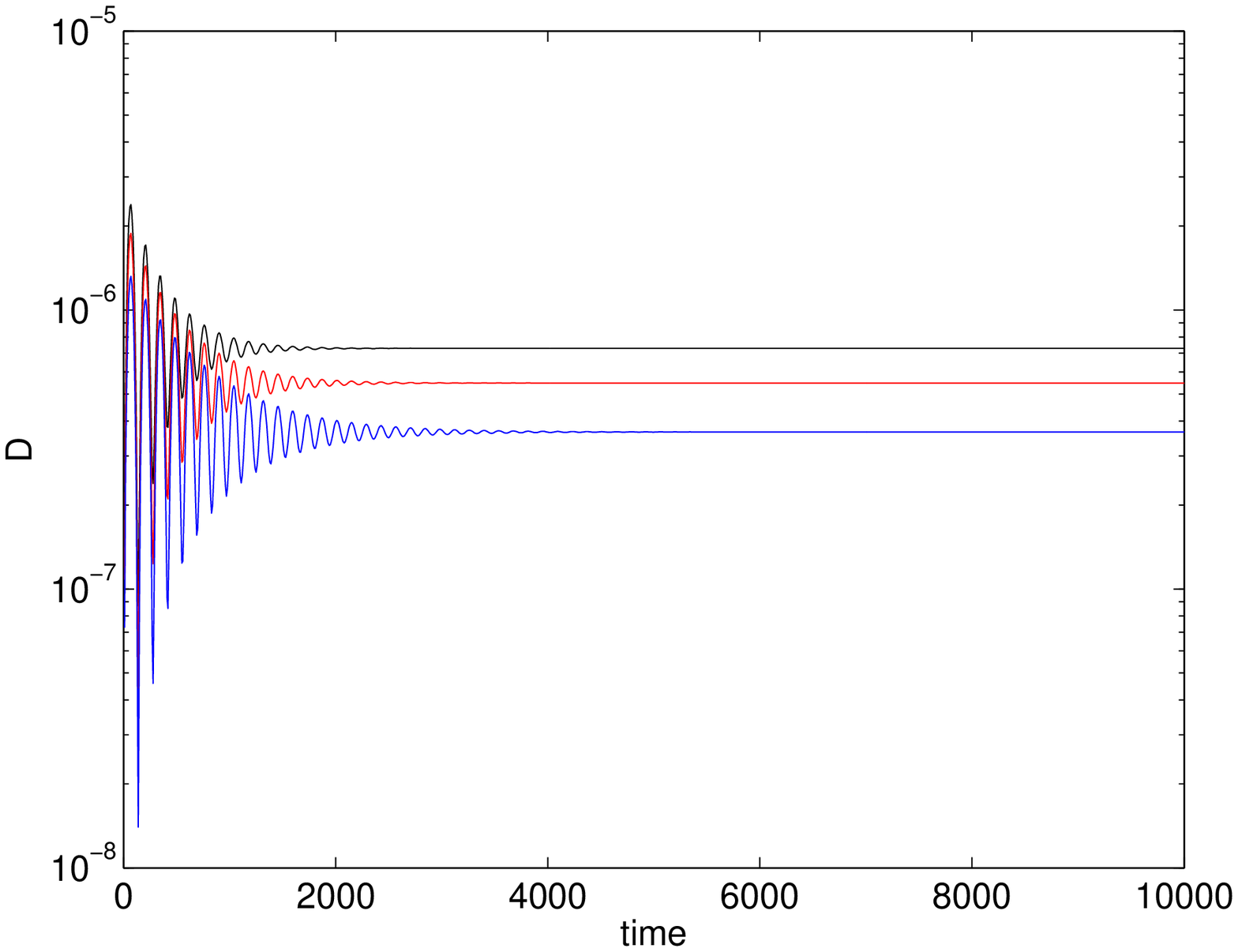}\label{fig:non9}}
\caption{The investigation of the Ekman number in the nonlinear regime. The dissipation $D$ versus time with $a=1\times10^{-3}$ and $k_x=k_y=k_z=1$. (a) The resonant frequency $\omega=-2/\sqrt{3}$. (b) $\omega=-1.16$ near the resonant frequency. (c) $\omega=-1.2$ far away from the resonant frequency. Black, red and blue lines denote the nonlinear regime at respectively $E=1\times10^{-3}$, $7.5\times10^{-4}$ and $5\times10^{-4}$.}\label{fig:Ekman}
\end{figure}

We then investigate the Ekman number. Usually viscosity has a stabilising effect (rigorously speaking, viscosity can be both stabilising and destabilising, e.g. in the parallel shear flow it can destabilise the inviscid flow with parabolic profile), and therefore, when the Ekman number decreases the nonlinear effect becomes significant. Figure \ref{fig:Ekman} shows the dissipation versus time for the three different Ekman numbers, i.e. $1\times10^{-3}$, $7.5\times10^{-4}$ and $5\times10^{-4}$, with the three frequencies as in figure \ref{fig:amplitude}. Figure \ref{fig:non7} shows that at the resonant frequency the lower Ekman number leads to the lower dissipation. This is \emph{opposite} to the prediction in the linear regime, namely the lower Ekman number leads to the higher dissipation at the resonant frequency as shown by the solid line in figure \ref{fig:lin3}. A tentative interpretation is that at the resonant frequency the lower Ekman number gives rise to the stronger nonlinear suppression for dissipation which wins out the linear enhancement for dissipation. Figure \ref{fig:non8} shows that near the resonant frequency the high and low Ekman numbers do not seem to have significant difference in respect of the time-averaged dissipation but the lower Ekman number leads to the stronger amplitude of oscillations of dissipation, which is presumably the instabilities caused by the stronger nonlinearity. Figure \ref{fig:non9} shows that far away from the resonant frequency the lower Ekman number leads to the lower dissipation. Moreover, the ratio of three dissipations at $E=1\times10^{-3}$, $7.5\times10^{-4}$ and $5\times10^{-4}$ is $1:0.75:0.5$, which is consistent with the prediction of $D\propto E$ in the linear regime as shown by the dash-dotted line in figure \ref{fig:lin3}. This again indicates that the nonlinear effect is negligible far away from the resonant frequency.

\section{Conclusion}\label{sec:conclusion}

In this work we study analytically and numerically the rotating flow driven by the harmonic force. In the linear regime we analytically derive the response to the harmonic force and the dissipation. The dissipation scales as $D\propto E^{-1}k^{-2}$ at the resonant frequency and $D\propto E$ at the other non-resonant frequencies. In the nonlinear regime we do the numerical calculations and compare to the linear regime. It is found that the effective dissipation ($D/a^2$) at the resonant frequency in the nonlinear regime is lower than in the linear regime and decreases with the increasing force amplitude, and however, this nonlinear suppression is negligible far away from the resonant frequency. Opposite to the linear regime, the lower Ekman number leads to the lower dissipation at the resonant frequency because of the stronger nonlinear effect. This nonlinear effect can be interpreted. At the resonant frequency, if the tidal force amplitude is large enough or the Ekman number is small enough, the nonlinear inertial force ($\bm u\cdot\bm\nabla\bm u$) takes its effect, such that the flow is suppressed, namely the flow intensity is weaker than without the nonlinear inertial force. Hence, the dissipation, which is equal to the enstrophy multiplied by viscosity, is also suppressed. Far away from the resonant frequency, the tidal response is weak and therefore the nonlinear effect is not striking.

In summary, when the frequency of the external harmonic force is close to the negative frequency of inertial wave in the unforced rotating flow, the dissipation can be greatly enhanced but the stronger nonlinear effect due to the stronger force amplitude or the lower Ekman number can suppress this enhancement, however, the dissipation at the frequency far away from the resonant frequency is small and the nonlinear effect is insignificant. Our numerical calculations about the nonlinear effect on the dynamical tide imply that the previous linear calculations \emph{overestimated} the tidal dissipation at the resonant frequency.

\section*{Acknowledgements}
This work was initiated in Princeton and completed in Shanghai. I am financially supported by the National Science Foundation’s Center for Magnetic Self-Organization under grant PHY-0821899 and the startup grant of Shanghai Jiaotong University.

\bibliographystyle{jfm}
\bibliography{paper}

\end{document}